\documentclass[%
 reprint,
nofootinbib,
 amsmath,amssymb,
 aps,
]{revtex4-2}

\usepackage{graphicx}

\usepackage{bm}

\usepackage{dcolumn}
\usepackage{booktabs} 
\usepackage{tabularx}
\usepackage{multirow}
\usepackage{array}

\usepackage{chemformula}

\usepackage{gensymb} 
\newenvironment{paperwidefigure}
  {\begin{figure*}}
  {\end{figure*}}

\usepackage{placeins}

\usepackage{siunitx}
\DeclareSIUnit\torr{torr}
\DeclareSIUnit\ppm{ppm}
\DeclareSIUnit\ppb{ppb}
\DeclareSIUnit\bar{bar}
\DeclareSIUnit\gauss{G}
\DeclareSIUnit\neV{\nano\electronvolt}
\DeclareSIUnit\peV{\pico\electronvolt}
\DeclareSIUnit\kms{km/s}
\usepackage[normalem]{ulem}
\usepackage{changes}

\usepackage{hyperref}
\usepackage{cleveref}

\crefname{section}{Sec.}{Secs.}
\crefname{figure}{Fig.}{Figs.}
\crefname{equation}{Eq.}{Eqs.}

\usepackage{xcolor}

\newcommand{\code}{\texttt}
\newcommand{\axionbloch}{\code{axionbloch}}

\begin{document}

\preprint{APS/123-QED}

\title{\axionbloch{}: an Open-Source Python Package for Simulating Axion-Induced Spin Dynamics}

\author{Yuzhe Zhang}
\email{Contact author: yuhzhang@uni-mainz.de}
\affiliation{Johannes Gutenberg-Universit{\"a}t Mainz, 55099 Mainz, Germany}
\affiliation{Helmholtz Institute Mainz, 55128 Mainz, Germany}
\affiliation{GSI Helmholtzzentrum für Schwerionenforschung GmbH, 64291 Darmstadt, Germany}

\begin{abstract}

The interaction of ultralight bosonic dark matter with spins can be interpreted as a pseudomagnetic field acting on normal matter.
Such interactions can be modeled as usual magnetic interactions using spin-evolution (Bloch) equations.
\axionbloch{}, an open-source Python package for simulating spin dynamics induced by both usual and exotic interactions, is presented. 
The numerical simulations serve as a tool for deriving axion signal signatures, which are crucial for designing experimental searches and data analysis. 
Simulations are calibrated against theoretical expectations, ensuring the accuracy of the simulated signals. 
\axionbloch{} is available at \url{https://github.com/Yuzhe98/AxionBloch}, allowing researchers to simulate pseudomagnetic signals under specific configurations of the axion models and experimental setups. 
The package is documented at \url{http://axionbloch.readthedocs.io/} and includes example scripts for application. 

\end{abstract}

\maketitle

\section{Introduction}

The axion was originally proposed to resolve the strong $CP$ problem in quantum chromodynamics\,\cite{PecceiQuinn1977_ConstraintsCP_PRD,Weinberg1978_NewLightBoson,PecceiQuinn1977_CPConservation_PRL,Wilczek1978_ProblemStrongPAndT}. 
Here, $C$ denotes charge conjugation and $P$ denotes parity. 
Later, more general axionlike particles (ALPs) were proposed in various extensions of the Standard Model associated with spontaneously broken global symmetries\,\cite{svrcek2006axions,Kim1987_LightPseudoscalars,JaeckelRingwald2010_ALPReview,Arvanitaki2010_StringAxiverse}. 
Concerning the cosmological implications, axions and ALPs are also well-motivated dark matter candidates\,\cite{Preskill1983_invisible_axion,Abbott1983_CosmologicalBoundAxion,Dine1983_not_so_harmless_axion,KhlopovMalomedZeldovich1985,Marsh2016_AxionCosmology}.
In this paper, axions and ALPs are referred to as ``axions'' for simplicity. 

Axions may possess couplings to ordinary Standard Model particles\,\cite{Graham2013_Newobservables}, including photons, gluons, and fermions. 
Crucially, axion field gradients ($\boldsymbol{\nabla}a$) may couple to fermionic spins through a pseudoscalar interaction. 
The Hamiltonian for this axion--fermion interaction is written as
\begin{equation}
    \mathcal{H}_{\mathrm{int}} = c g_\mathrm{aNN} \boldsymbol{\nabla}a \cdot \boldsymbol{S}\,,
\end{equation}
where $g_\mathrm{aNN}$ is the axion--fermion coupling strength, $c$ is the speed of light, and $\boldsymbol{S}$ denotes the physical spin operator, $\boldsymbol{S}=\hbar\boldsymbol{\sigma}/2$, with $\boldsymbol{\sigma}$ the vector of Pauli matrices. 
The axion field gradient can be regarded as a pseudomagnetic field that mimics a real magnetic field. The corresponding Hamiltonian can be written as
\begin{equation}
    \mathcal{H} = -\gamma \boldsymbol{B} \cdot \boldsymbol{S} \,,
\end{equation} 
where $\gamma$ is the gyromagnetic ratio of the spin, and $\boldsymbol{B}$ is the magnetic field. 
Comparing the two Hamiltonians shows that the gradient of the axion field acts effectively as a magnetic field. 
Therefore, the concept of an axion-induced pseudomagnetic field is introduced to represent this effective magnetic field, which is defined as
\begin{equation}
    \boldsymbol{B}_a = -\frac{c g_\mathrm{aNN}}{\gamma}\boldsymbol{\nabla}a\,.
\end{equation}
The precise form of the pseudomagnetic field depends on the axion model.

The pseudomagnetic field induces spin dynamics, providing a pathway for detection in precision measurement experiments such as nuclear magnetic resonance (NMR) and comagnetometer-based searches\,\cite{ouellet2019first, du2018search, ALPHA2023,aja2022canfranc,CAPP2024,brouwer2022projected,alesini2023future,ZhongL2018results,garcia2024searchaxiondarkmatter,Quiskamp2024_ORGAN,rettaroli2024search,Gramolin2021_Ferromagnetic,gavilan2024searching,bloch2105nasduck,abel2023_HgDM,karanth2023first,LeeJunyi2023_NeutronSpin,capozzi2023new, CASPErProposal2014, kimball2018overview, wang2018application, aybas2021_SolidStateNMR, CASPEr_ZULF_2019,Walter2025_CASPErG}. 
There have been dedicated studies of the axion-induced spin dynamics\,\cite{Aybas_2021_Quantum_sensitivity, Dror2023_SensitivityAxion, Yuzhe2023_FrequencyScanning}, providing axion signal signatures and experimental sensitivity to axion fields. 
However, to our knowledge, there have not been studies taking advantage of numerical methods to verify the derived axion signal signatures. 
On the other hand, as more axion models and experiments are being proposed, such as cosmic axion background\,\cite{Dror2021_CosmicAxionBackground,Nitta2023_ADMX_CosmicAxionBackground}, fine-grained axion\,\cite{OHare2025_Finegrained}, and quadratically coupled axion\,\cite{GanXucheng2026_quadraticPTA, huang2026earthmatterenhancedaxion}, a tool is needed to systematically simulate the axion-induced spin dynamics under various physical conditions and axion field configurations.

In this work, \axionbloch{}, an open-source package designed to simulate spin dynamics induced by axion fields, is introduced. 
The interface is implemented in Python, a programming language known for its simplicity and readability, while the underlying numerical integration is implemented in \code{C++} for efficiency. 
The package provides a flexible framework for modeling axion fields and experimental configurations. 
By numerically solving spin-evolution equations, \axionbloch{} enables the prediction and analysis of experimentally observable signals.

\section{Capabilities}
\label{sec:capabilities}

In this section, the main components of \axionbloch{} are described. The section starts with the units, physical quantities, and constants (\cref{sec:units_quantities_constants}). Then, it continues with the physics and the algorithm for the numerical simulation (\cref{sec:algorithm}), followed by calibrations (\cref{sec:calibration}), comparisons with NMR measurements (\cref{sec:nmr_simulation}), and an example simulation of axion-induced signals (\cref{sec:simulation}). 

\subsection{Units, physical quantities, and constants}
\label{sec:units_quantities_constants}

\begin{figure}
\center
\fbox{\includegraphics[width=0.45\textwidth,clip]{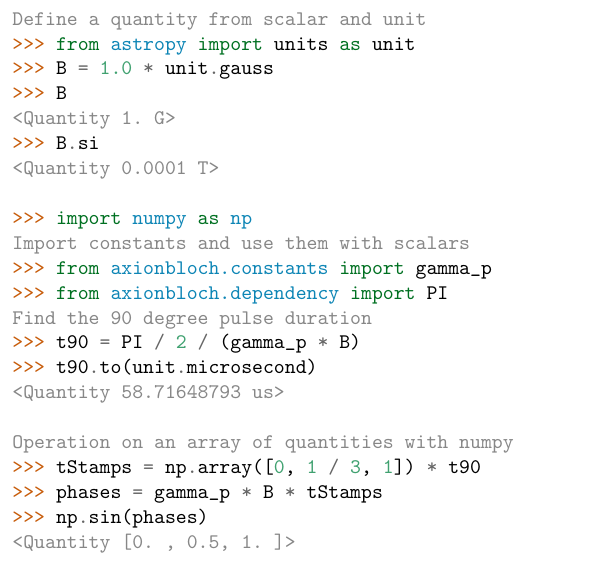}}
\caption{Examples of operations on physical quantities using \texttt{astropy.units} and \texttt{axionbloch.constants}. }
\label{code:quantities}
\end{figure}

The package takes advantage of \texttt{astropy}\footnote{\url{https://astropy.org/}}\,\cite{Astropy2013} to deal with units and physical quantities. 
Using \texttt{astropy.units}, physical quantities are created by attaching units to Python scalars or \texttt{NumPy}\footnote{\url{https://numpy.org/}}\,\cite{Harris2020_NumPy} arrays. 
Quantities can be converted between dimensionally equivalent units using the \texttt{to()} method.
In \texttt{astropy}, unit definitions are included in both the International System of Units (SI) and the Centimeter-Gram-Second (CGS) systems. 
Values in SI or CGS units can be found in the quantity's properties \texttt{.si} and \texttt{.cgs}.
A code example can be found in \cref{code:quantities}. 

In \texttt{axionbloch.constants}, a few commonly-used constants in axion and NMR research, such as gyromagnetic ratios, are defined.
The constants are instances of \texttt{astropy.units.Quantity}; therefore, all operations on physical quantities are supported for the constants. 

\subsection{Algorithm}
\label{sec:algorithm}

\subsubsection{Spin dynamics}
\label{sec:spindynamics}

Considering the current experiments looking for axion-induced spin dynamics, \axionbloch{} focuses on spin-$1/2$ magnetic resonance. 
Different spins may be addressed in future updates. 
Magnetization is used as the observable to characterize the spin ensemble. 
In the presence of magnetic fields or axion-induced pseudomagnetic fields, the magnetization $\boldsymbol{M}$ evolves as
\begin{align}
\frac{d M_x}{dt} &= \gamma \left( \boldsymbol{M} \times \boldsymbol{B}_{\mathrm{eff}} \right)_x - \frac{M_x}{T_2}\,,\label{eq:bloch_x} \\
\frac{d M_y}{dt} &= \gamma \left( \boldsymbol{M} \times \boldsymbol{B}_{\mathrm{eff}} \right)_y - \frac{M_y}{T_2}\,,\label{eq:bloch_y} \\
\frac{d M_z}{dt} &= \gamma \left( \boldsymbol{M} \times \boldsymbol{B}_{\mathrm{eff}} \right)_z - \frac{M_z - M_0}{T_1}\,,\label{eq:bloch_z}
\end{align}
where $\gamma$ is the gyromagnetic ratio, $\boldsymbol{B}_{\mathrm{eff}}$ is the total effective field, $T_1$ and $T_2$ are relaxation times, and the subscripts $x$, $y$, $z$ denote the Cartesian components. 
The bias field is usually much stronger than the pseudomagnetic field. 
Therefore, the rotating coordinate frame (RCF) is adopted to remove the Larmor precession due to the bias field, allowing efficient simulation of the slower axion-induced dynamics of interest. 

The transverse relaxation time, characterized by time constant $T_2^*$, includes the $T_2$ relaxation and the dephasing due to the inhomogeneity of the bias field. 
Such inhomogeneity of the bias field is simulated by sampling the spread of the bias fields in the \texttt{Magnet} instance, and solving the Bloch equations for individual bias fields. 

\subsubsection{Numerical integration}
\label{sec:numericalintegration}

To derive the time evolution of the magnetization, \axionbloch{} numerically integrates the Bloch equations using the fourth-order Runge--Kutta (RK4) method. 
Given a time step $\Delta t$, an effective magnetic field, and initial magnetization $\boldsymbol{M}^0$, the magnetization at the $(n+1)$-th step can be found via 
\begin{equation}
\boldsymbol{M}^{n+1}
=
\boldsymbol{M}^n
+
\frac{\Delta t}{6}
\left(
\boldsymbol{k}_1 + 2\boldsymbol{k}_2 + 2\boldsymbol{k}_3 + \boldsymbol{k}_4
\right)\,.
\end{equation}
Here,
\begin{align}
\boldsymbol{k}_1 &= \left.\frac{d\boldsymbol{M}}{dt}\right|_{\boldsymbol{M}=\boldsymbol{M}^n}\,, \\
\boldsymbol{k}_2 &= \left.\frac{d\boldsymbol{M}}{dt}\right|_{\boldsymbol{M}=\boldsymbol{M}^n + \frac{1}{2}\boldsymbol{k}_1\Delta t}\,, \\
\boldsymbol{k}_3 &= \left.\frac{d\boldsymbol{M}}{dt}\right|_{\boldsymbol{M}=\boldsymbol{M}^n + \frac{1}{2}\boldsymbol{k}_2\Delta t}\,, \\
\boldsymbol{k}_4 &= \left.\frac{d\boldsymbol{M}}{dt}\right|_{\boldsymbol{M}=\boldsymbol{M}^n + \boldsymbol{k}_3\Delta t}\,,
\end{align}
where $d\boldsymbol{M}/dt$ can be found using the Bloch equations [Eqs.\,(\ref{eq:bloch_x}--\ref{eq:bloch_z})]. 
This fourth-order method yields a local truncation error of $\mathcal{O}(\Delta t^5)$. 
The time step $\Delta t$ is usually set to be at least one order of magnitude smaller than the characteristic timescales of the system, such as relaxation times. 

\subsubsection{Implementation}
\label{sec:implementation}

Using object-oriented programming, the algorithm is implemented in a modular way to separate experimental configuration, axion field modeling, and numerical integration, allowing for flexible configuration of simulations. 
A diagram of the architecture of \axionbloch{} is shown in \cref{fig:architecture}. 
\begin{paperwidefigure}[htb]
    \centering
    \fbox{\includegraphics[width=0.8\textwidth]{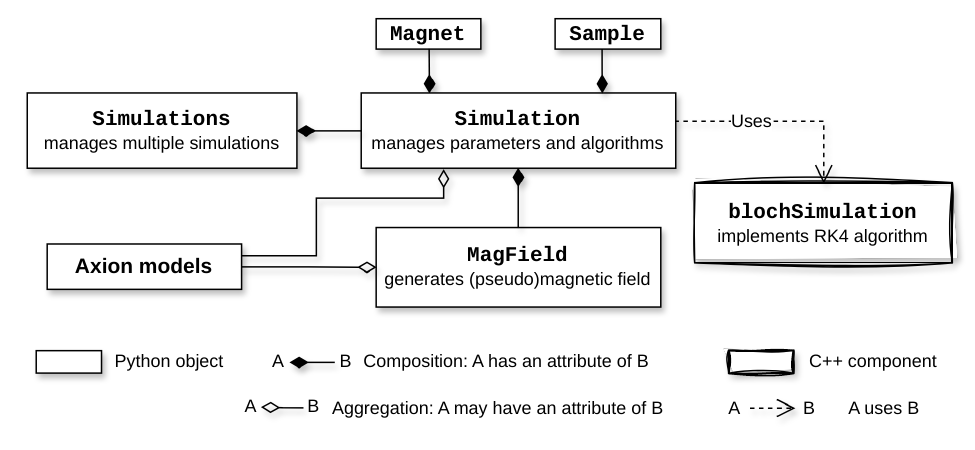}}
    \caption{Architecture of \axionbloch{}. Each \texttt{Simulation} instance imports parameters from \texttt{Magnet}, \texttt{Sample}, and \texttt{MagField} instances. For axion simulations, axion-model objects are required both for \texttt{Simulation} to configure the simulation and for \texttt{MagField} to generate the pseudomagnetic field. \texttt{Simulations} manages one or multiple \texttt{Simulation} instances. Numerical integration is performed via the \texttt{blochSimulation} API, which provides Python access to the underlying \code{C++} implementation. 
    }
    \label{fig:architecture}
\end{paperwidefigure}

On the Python side, classes \texttt{Sample}, \texttt{Magnet}, and \texttt{MagField} encapsulate the relevant experimental parameters, including sample relaxation times, magnetic field strength, and inhomogeneity. 
The axion models are implemented by dedicated classes, such as \texttt{MilkyWayAxionHalo}, which carry the properties of the axion fields. Together with the method \texttt{MagField.setAxionFields()}, the pseudomagnetic fields can be generated. 
Different axion models can be incorporated as separate classes without modifying the simulation framework. 
Simulation is handled within the \texttt{axionbloch.SimuTools} module. 
In particular, the classes \texttt{Simulation} and \texttt{Simulations} manage the interaction between experimental configurations and axion models, organizing the parameters and execution of single or multiple simulation runs. 

The RK4 algorithm is implemented in a dedicated \code{C++} library. 
The application programming interface (API) for this library is \texttt{blochSimulation}. 
This \code{C++} backend is interfaced with Python using \texttt{pybind11},\footnote{\url{http://pybind11.readthedocs.io/}} enabling cooperation between high-level Python workflow management and low-level efficient numerical computation.

\subsection{Calibration}
\label{sec:calibration}

Accurate calibration is essential for ensuring that simulated signals reflect the underlying physical system and can be compared with experimental observables. 
In particular, calibration examines how NMR signals depend on model parameters, including magnetic fields and relaxation times ($T_1$, $T_2$, and $T_2^*$). 
It also helps verify the correctness and stability of the numerical implementation. 
This section presents calibration simulations alongside theoretical predictions. 

\subsubsection{Methodology}
\label{sec:calibration:methodology}

The simulation accuracy is quantified by the relative $L^2$ error, $\epsilon_{L^2}$, between the simulated and theoretically predicted magnetization:
\begin{equation}
    \epsilon_{L^2}
    =
    \frac{
        \left[
        \sum_{i\in\mathcal{V}}
        \left|\boldsymbol{M}_{i}^\mathrm{simu}-\boldsymbol{M}_{i}^\mathrm{theo}\right|^2
        \right]^{1/2}
    }{
        \left[
        \sum_{i\in\mathcal{V}}
        \left|\boldsymbol{M}_{i}^\mathrm{theo}\right|^2
        \right]^{1/2}
    }\,,
    \label{eq:relative_l2_error}
\end{equation}
where $\boldsymbol{M}_{i}=(M_{x,i},\,M_{y,i},\,M_{z,i})$ is the magnetization at the $i$-th sample, and $\mathcal{V}$ is the set of samples included in the comparison. 
Ideally, $\epsilon_{L^2}$ should be zero, indicating a perfect match between the simulated and theoretical results. 
However, simulations advance in discrete steps. 
Because the integration step size is finite, the simulated magnetization does not exactly follow its continuous-time evolution, resulting in a nonzero $\epsilon_{L^2}$. 
Using smaller integration steps, \textit{i.e.}, higher simulation rates, should reduce the error at the cost of increased computational time. 
At sufficiently high simulation rates, other factors, such as floating-point precision, should dominate the error. 
Convergence tests are performed to ensure that the accuracy of the simulation results improves with higher simulation rates.

The complete calibration scripts are included in the \axionbloch{} test suite. 
The following examples illustrate the calibration methodology.
Unless otherwise stated, the examples use a thermally polarized proton-NMR sample and an axial bias field, $\boldsymbol{B_{0}} = B_0 \boldsymbol{\hat{e}}_z$. 
The Larmor frequencies of the examples are set to \SI{1}{\MHz}. 

\subsubsection{Excitation}
\label{sec:calibration:excitation}

In NMR experiments, magnetic-field excitations are applied to tip the magnetization vector.
Like in experiments, simulation excitations must be calibrated so that the magnetization is tipped by the intended angle. 
In this calibration, excitations are applied in the $x$-$y$ plane, and the magnetization is expected to tip by an angle
\begin{equation}
    \theta = \gamma B_{xy} \Delta t_{xy}\,,
\end{equation}
where $B_{xy}$ is the excitation amplitude and $\Delta t_{xy}$ is the excitation duration.
Theoretically, the corresponding transverse magnetization amplitude is 
\begin{equation}
    M_{xy}^\mathrm{theo} = M_\mathrm{eqb} \sin\theta\,
\end{equation}
after the excitation. 
Here $M_\mathrm{eqb}$ is the equilibrium magnetization.
The tip angle is swept from \SI{0}{\degree} to \SI{360}{\degree}, and the discrepancy between the simulated and theoretical transverse magnetizations is quantified using $\epsilon_{L^2}$ from \cref{eq:relative_l2_error}.
An example of Rabi oscillation (transverse magnetization versus user-specified tip angle) is illustrated in \cref{fig:calibration-excitation}\,(a). 

\begin{paperwidefigure}[htb]
    \centering
    \includegraphics[width=0.9\textwidth]{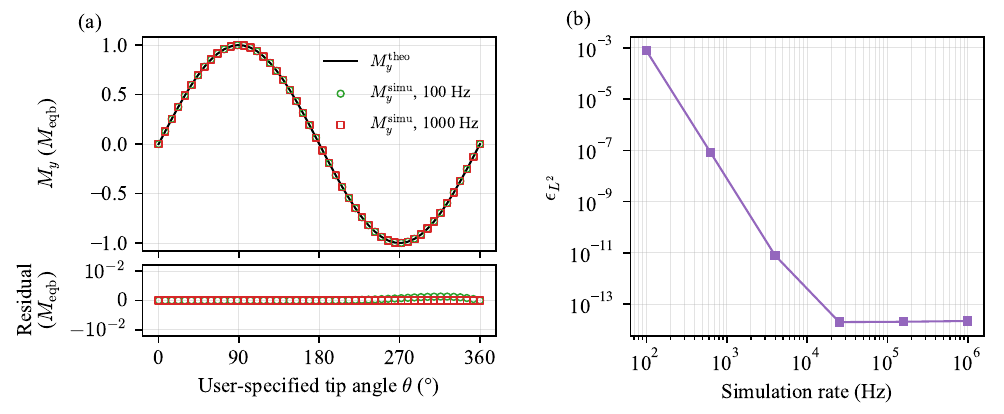}
    \caption{
    Calibration of magnetic-field excitations. (a) Simulated transverse magnetization $M_y$ after a \SI{100}{\milli\second} magnetic-field excitation along the $x$ axis, plotted as a function of tip angle. The equilibrium magnetization $M_\mathrm{eqb}$ is used as the reference value. The tip angle ranges from \SI{0}{\degree} to \SI{360}{\degree}. Two simulation rates (\SI{100}{\Hz} and \SI{1000}{\Hz}) illustrate the effect of simulation rate on accuracy. Simulation results ($M^\mathrm{simu}$, markers) are compared with the theoretical prediction ($M^\mathrm{theo}$, solid black line). The lower plot shows the residuals, defined as the differences between the simulated and theoretical values. (b) Relative $L^2$ error $\epsilon_{L^2}$ versus simulation rate.} 
    \label{fig:calibration-excitation}  
\end{paperwidefigure}

As can be seen in \cref{fig:calibration-excitation}\,(b), $\epsilon_{L^2}$ decreases as the simulation rate increases, then plateaus when other factors dominate, as expected for a convergent numerical method. 
The error is generally below \SI{1}{\percent} at a moderate simulation rate of \SI{100}{\Hz}, indicating that the simulation is sufficiently accurate and efficient for the applications considered here.

\subsubsection{Relaxations}
\label{sec:calibration:relaxation}

For $T_1$ relaxation, the magnetization approaches its equilibrium value according to
\begin{equation}
    M_z^\mathrm{theo}(t)
    = M_\mathrm{eqb}
    + \left(M_z(0)-M_\mathrm{eqb}\right)e^{-t/T_1}\,.
\end{equation}
The simulation examples presented here use two initial conditions, $M_z(0)=0$ and $M_z(0)=10\,M_\mathrm{eqb}$, and track the magnetization as it relaxes toward $M_\mathrm{eqb}$. 
As can be seen in \cref{fig:calibration-t1}\,(a), the simulated $M_z$ is compared with the theoretical prediction, and the residual is shown below. 
In \cref{fig:calibration-t1}\,(b), the relative $L^2$ error $\epsilon_{L^2}$ between the simulated and theoretical magnetization is plotted against the simulation rate. 
The error decreases as the simulation rate increases, until it plateaus at a level determined by floating-point precision.

\begin{paperwidefigure}[htb]
    \centering
    \includegraphics[width=0.9\textwidth]{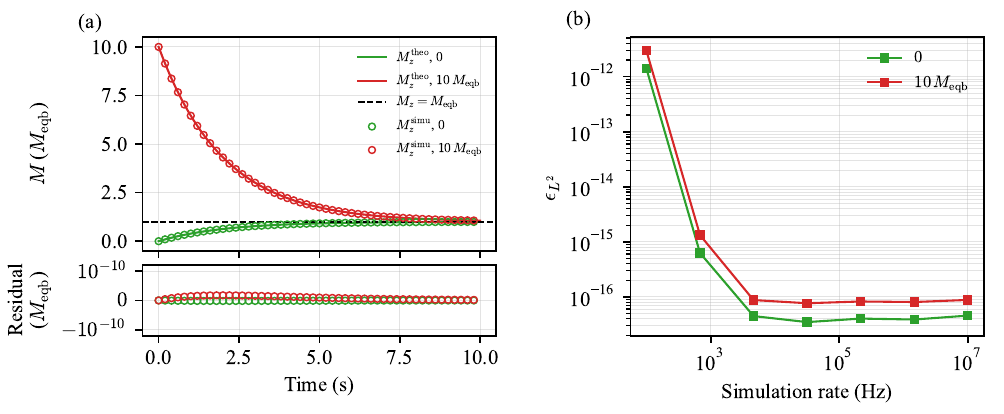}
    \caption{
    Examples of $T_1$ relaxation and simulation-rate convergence. (a) Longitudinal magnetization $M_z$ as a function of time. The initial magnetization is either $0$ or hyperpolarized to $10\,M_\mathrm{eqb}$ and relaxes toward $M_\mathrm{eqb}$ with $T_1=\SI{2}{\second}$. Simulation results ($M^\mathrm{simu}$, markers, simulation rate \SI{100}{\Hz}) are compared with the theoretical predictions ($M^\mathrm{theo}$, solid lines); residuals are shown below. (b) Relative $L^2$ error $\epsilon_{L^2}$ versus simulation rate for both initial conditions.}
    \label{fig:calibration-t1}
\end{paperwidefigure}

For $T_2$ relaxation, the magnetization starts along the $y$ axis with $M_y=M_\mathrm{eqb}$ and exponentially decays toward zero according to
\begin{equation}
    M_y^\mathrm{theo}(t) = M_\mathrm{eqb}e^{-t/T_2}\,.
\end{equation} 
In \cref{fig:calibration-t2}\,(a), simulations of $M_y$ at \SI{100}{\Hz} and \SI{1000}{\Hz} are compared with the theoretical prediction; the residuals are shown below. 

\begin{paperwidefigure}[htb]
    \centering
    \includegraphics[width=0.9\textwidth]{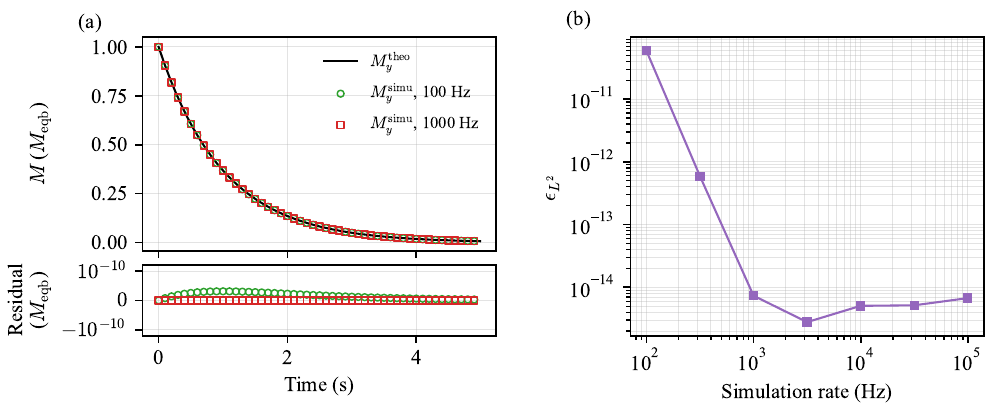}
    \caption{
    Examples of $T_2$ relaxation and simulation-rate convergence. (a) Transverse magnetization $M_y$ as a function of time. The magnetization starts at $M_y=M_\mathrm{eqb}$ and decays toward zero with $T_2=\SI{1}{\second}$. Simulation results ($M^\mathrm{simu}$, markers, at \SI{100}{\Hz} and \SI{1000}{\Hz}) are compared with the theoretical prediction ($M^\mathrm{theo}$, solid black line); residuals are shown below. (b) Relative $L^2$ error $\epsilon_{L^2}$ versus simulation rate.}
    \label{fig:calibration-t2}
\end{paperwidefigure}

For both relaxation calibrations, $\epsilon_{L^2}$ is calculated from the difference between the simulated and theoretical magnetizations using \cref{eq:relative_l2_error}.
The error decreases as the simulation rate increases, then plateaus at a level determined by floating-point precision, as shown in \cref{fig:calibration-t1}\,(b) and \cref{fig:calibration-t2}\,(b).
In general, the error is well below \SI{1}{\percent}, indicating a good level of accuracy for the simulations.

\subsection{NMR simulation}
\label{sec:nmr_simulation}

\subsubsection{Pulsed NMR: free decay and spin echo}
\label{sec:pNMR}
In the pulsed-NMR simulation, one \SI{90}{\degree} pulse and two \SI{180}{\degree} pulses are applied. 
The magnetization is expected to exhibit the following behavior\,\cite{CPMG, Abragam1961_NuclearMagnetism}:
\begin{itemize}
    \item oscillate at the Larmor frequency;
    \item be tipped by $\SI{90}{\degree}$ to the x-y plane by the $\SI{90}{\degree}$ pulse;
    \item free decay with relaxation time $T_2^*$ in the absence of pulses;
    \item refocus (echo) after $\SI{180}{\degree}$ pulses; 
    \item show decaying amplitude of echoes with relaxation time $T_2$.
\end{itemize}
The behaviors described above can be found in the calibration example illustrated in \cref{fig:calibration_T2_T2star}\,(a), indicating correct simulation behavior. 

\begin{paperwidefigure}[htb]
    \centering
    \includegraphics[width=0.9\textwidth]{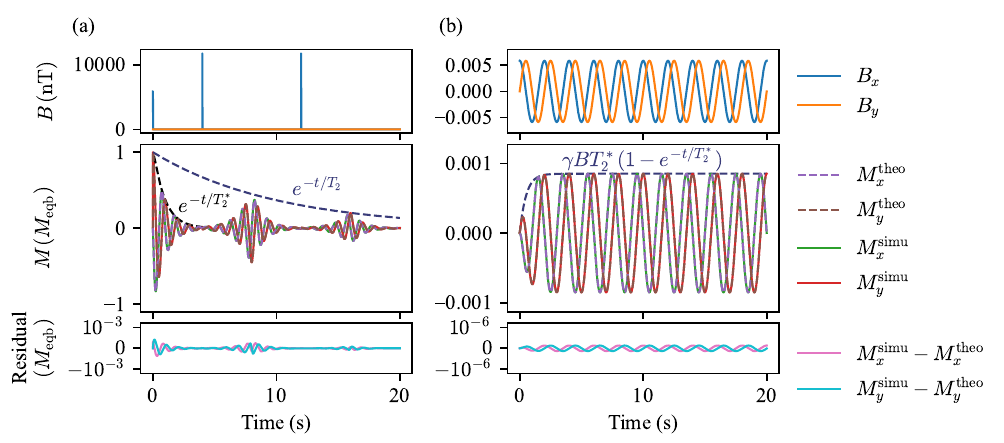}
    \caption{Simulated magnetic fields along the $x$ and $y$ directions (upper row), the resulting NMR signals (middle row), and the corresponding residuals (lower row). (a) After $\SI{90}{\degree}$ and $\SI{180}{\degree}$ magnetic-field pulses in $x$-$y$ plane, free decay (relaxation time $T_2^*$) and echo signals (relaxation time $T_2$) are generated. (b) Excited by a weak CW magnetic field, the NMR signals increase with $\gamma B T_2^* (1 - e^{-t/T_2^*})$, where $B = \sqrt{B_x^2+B_y^2}$ is the magnitude of the excitation.
    }
    \label{fig:calibration_T2_T2star}
\end{paperwidefigure}

\subsubsection{Continuous-wave NMR}
\label{sec:CW_NMR}
In the continuous-wave (CW) NMR simulation, an oscillating magnetic field is applied. 
When the field is weak ($\gamma B T_2^*\ll 1$) and on resonance, the transverse magnetization normalized to its equilibrium value grows as $\gamma B T_2^* (1-e^{-t/T_2^*})$\,\cite{Abragam1961_NuclearMagnetism}.
An example is shown in \cref{fig:calibration_T2_T2star}\,(b), and the simulation follows the expected behavior.

\subsection{Axion simulation}
\label{sec:simulation}

Similar to the NMR examples, axion-induced NMR signals are generated by solving the Bloch equations with the axion-induced pseudomagnetic field as part of the effective magnetic field. 
An example of a commonly considered model -- axion constituting the Milky Way galactic dark matter halo\,\cite{Evans2019_SHMRefined,Gramolin2022Feb_SpectralSignatures} is implemented in the class \texttt{MilkyWayAxionHalo}. 
In this model, the axion field is treated as an oscillating field with a frequency determined by the axion mass, and a linewidth (characterizing the stochasticity) determined by the velocity dispersion of the virialized dark matter halo. 
The sample is chosen to be \ch{^{129}Xe}, polarized to $\SI{50}{\percent}$ initially, and the bias field is applied along the z direction. 
The details of the simulation configuration can be found in the code example in \cref{code:axion_example}. 
Readers are referred to \cref{fig:architecture} which illustrates the relationship between the objects in the code example. 
The simulated pseudomagnetic fields and the amplitude of the induced transverse magnetization are shown in \cref{fig:MilkyWayAxionHalo}, in which the signal increases with time at the beginning of the simulation due to the pseudomagnetic field, and then decays due to the $T_1$ relaxation. 
The efficiency of the simulation is limited by time consumption for the numerical integration of the Bloch equations. 
The total number of integration steps $N$ can be found by
\begin{align}
    N &= \text{\texttt{Simulation.rate}} \times \text{\texttt{Simulation.duration}} \nonumber\\
    &\quad \times \text{\texttt{MagField.numFields}} \times \text{\texttt{Magnet.numPt}}\,,
\end{align}
where \texttt{Simulation.rate} and \texttt{Simulation.duration} are the simulation rate and duration, \texttt{MagField.numFields} is the number of random pseudomagnetic-field realizations used to account for axion-field stochasticity, and \texttt{Magnet.numPt} is the number of bias-field values sampled to account for bias-field inhomogeneity. 

To accelerate the simulation, the independent realizations of the (pseudo)magnetic fields are parallelized using Open Multi-Processing (OpenMP)\,\cite{OpenMP2024} in \texttt{blochSimulation}.
The number of OpenMP threads is controlled by the runtime environment and may therefore vary with the user's hardware and OpenMP settings.
Within each realization, the magnetization-packet and time-stepping loops are evaluated serially. 

For the example in \cref{code:axion_example}, the total number of integration steps is \SI{7e8}{}, and the runtime is \SI{2.7}{\s} on a personal computer with an Intel\textsuperscript{\textregistered} Core\textsuperscript{\texttrademark} Ultra 7 155H CPU (base frequency: \SI{1.40}{\GHz}). 
This runtime allows users to explore multiple parameter regimes through repeated simulations. 

\begin{paperwidefigure}
\center
\fbox{
  \includegraphics[
    height=0.9\textheight,
    keepaspectratio,
    clip
  ]{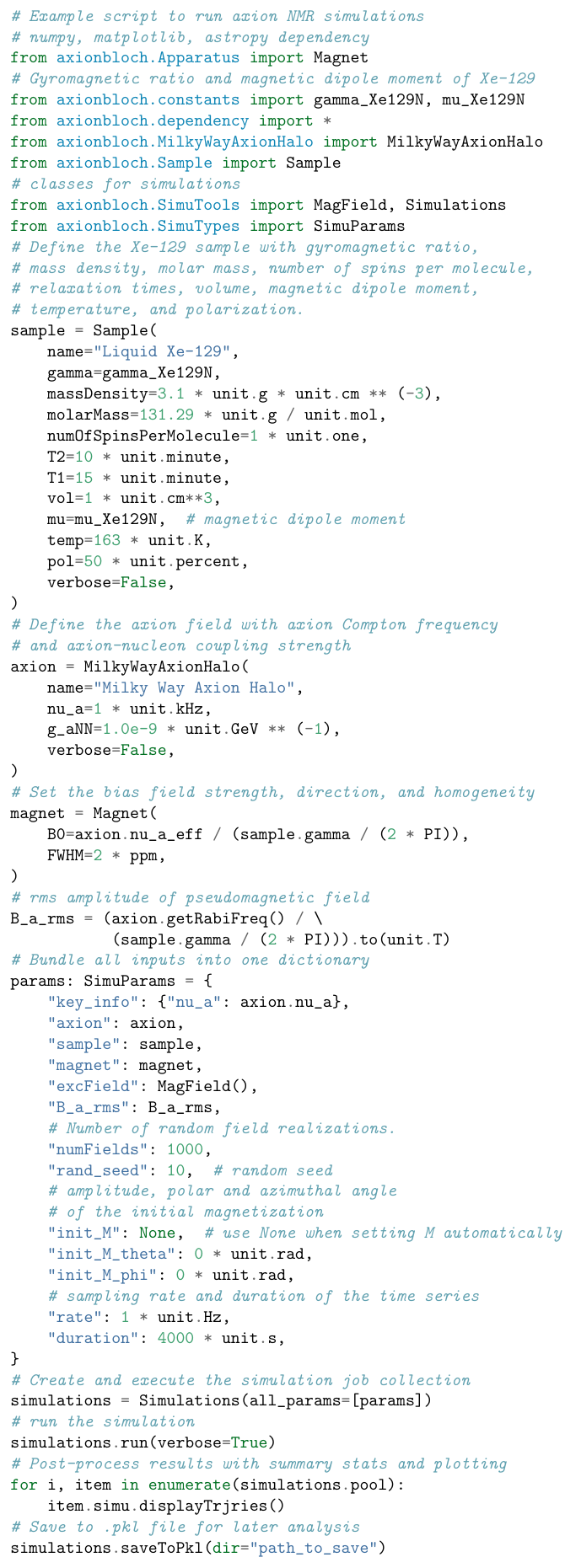}%
}
\caption{Example code of an axion simulation.}
\label{code:axion_example}
\end{paperwidefigure}

\begin{paperwidefigure}[htb]
    \centering
    \includegraphics[width=0.9\textwidth]{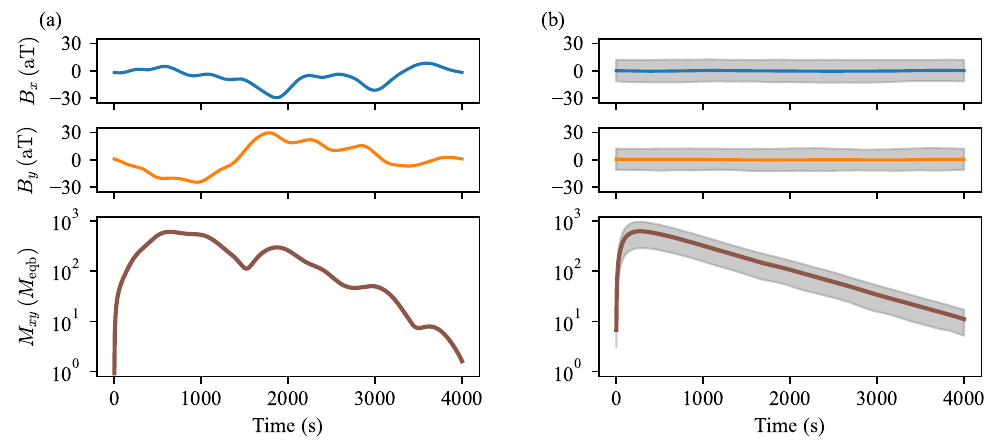}
    \caption{Simulations of local virialized axion dark matter halo. From top to bottom: pseudomagnetic fields in the transverse plane ($B_x$ and $B_y$) and the magnitude of the transverse magnetization $M_{xy}=\sqrt{M_x^2 + M_y^2}$. Left: one single simulation. Right: 1000 simulations to account for the stochasticity. The solid lines indicate the value (left column) or mean value (right column) of $B_x$, $B_y$, and $M_{xy}$, while the gray shadows indicate the $\pm \sigma$ (standard deviation) range of the 1000 simulations. The sample for the simulations is assumed to be polarized to $\SI{50}{\percent}$ initially.
    }
    \label{fig:MilkyWayAxionHalo}
\end{paperwidefigure}

\section{Development}
\label{sec:dev-approach}

Development of \axionbloch{} is managed using \texttt{git},\footnote{\url{http://git-scm.com}} with the code hosted on a public GitHub repository.\footnote{\url{https://github.com/Yuzhe98/AxionBloch}} 
Changes to the main codebase follow a pull-request workflow, allowing contributions to be reviewed before they are merged. Users can participate by forking the repository, making changes in their fork, and submitting pull requests for review by the developers.
As \axionbloch{} is intended as a collaborative resource, user contributions and suggestions are encouraged and appreciated. 

\axionbloch{} includes a suite of automated tests to check correctness and reliability. 
The tests use the \texttt{pytest} framework,\footnote{\url{https://docs.pytest.org/}} which supports calibration of physical processes and testing of functions, classes, and modules. 
The test suite covers core functionalities such as:
\begin{itemize}
    \item initialization and manipulation of classes; 
    \item numerical integration of the Bloch equations;
    \item calibrations;
    \item handling of axion field configurations.
\end{itemize}
Tests can be executed locally. 
In the future, tools may be added for online testing considering the demand of testing. 

Documentation for \axionbloch{} is available on Read the Docs.\footnote{\url{http://axionbloch.readthedocs.io}}
It is built using \texttt{Sphinx}\footnote{\url{http://www.sphinx-doc.org}} and explains the API and provides instructions for installation, configuration, and use.

As new axion models and experiments emerge, \axionbloch{} may be extended to support them and to help characterize the resulting spin dynamics. 
Future versions may also support zero-to-ultra-low-field (ZULF) NMR simulations. 

\section{Conclusions}
\label{sec:conclusions}

\axionbloch{} has been presented as a framework for simulating spin dynamics induced by magnetic or axion fields under a variety of physical conditions. 
The package adopts SI units so that the inputs and outputs are directly comparable to experimental parameters and observables. 
Numerical integration of the Bloch equations is performed using a fourth-order Runge--Kutta method in the rotating coordinate frame, enabling stable and efficient simulation. 
Comprehensive calibrations have been performed to ensure the accuracy of the simulations.
An example simulation of axion-induced signals from a local virialized axion dark matter halo is presented to demonstrate the package's capabilities. 

As an open-source, documented, and tested tool, \axionbloch{} aims to facilitate studies of axion-induced spin dynamics in precision-measurement experiments and to provide a platform for method development and the incorporation of new axion models.

Note that this package can be of interest to the NMR community for its applications in, for example, educational purposes. 
For example, students can specify pulse sequences for the simulation and see the visualized simulation results.

\section*{Data Availability}
\label{sec:data-availability}
The data and code that generate this manuscript are available at \url{
https://github.com/Yuzhe98/AxionBloch-paper}. 

\begin{acknowledgments}

The author acknowledges helpful discussions with Hendrik Bekker, Dmitry Budker, and Alexander Sushkov, and thanks the CASPEr Collaboration for its support. 
The author is grateful to the communities supporting the open-source software and tools: \texttt{git}, \texttt{NumPy}, 
\texttt{SciPy}\footnote{\url{https://docs.scipy.org/}}\,\cite{Virtanen2020_SciPy}, 
\texttt{matplotlib}\footnote{\url{https://matplotlib.org/}}, 
\texttt{astropy}, 
\texttt{pybind11}, 
\texttt{IPython}\footnote{\url{https://ipython.org/}}\,\cite{Perez2007_IPython}, 
\texttt{Jupyter}\footnote{\url{https://jupyter.org/}}\,\cite{Granger2021_Jupyter}, 
\texttt{pytest}, 
\texttt{Sphinx}, 
GitHub\footnote{\url{https://github.com/}} 
and 
Read the Docs\footnote{\url{https://about.readthedocs.com/}}. 
This work has been supported by the Cluster of Excellence ``Precision Physics, Fundamental Interactions, and Structure of Matter'' (PRISMA++ EXC 2118/2) funded by the
German Research Foundation (DFG) within the German Excellence Strategy (Project ID 390831469).

\end{acknowledgments}

\bibliographystyle{unsrt}
\bibliography{references}

\end{document}